\documentclass[prb, twocolumn, superscriptaddress]{revtex4}
\usepackage{graphicx}

\begin{document}

\title{Introducing single Mn$^{2+}$ ions into spontaneously coupled quantum dot pairs.} 

\author{M. Koperski}
\email{Maciej.Koperski@fuw.edu.pl}
\affiliation{Institute of Experimental Physics, University of Warsaw, Ho\.{z}a 69, 00-681 Warsaw, Poland}
\author{M. Goryca}
\affiliation{Institute of Experimental Physics, University of Warsaw, Ho\.{z}a 69, 00-681 Warsaw, Poland}
\author{T. Kazimierczuk}
\affiliation{Institute of Experimental Physics, University of Warsaw, Ho\.{z}a 69, 00-681 Warsaw, Poland}
\author{T. Smole\'{n}ski}
\affiliation{Institute of Experimental Physics, University of Warsaw, Ho\.{z}a 69, 00-681 Warsaw, Poland}
\author{A. Golnik}
\affiliation{Institute of Experimental Physics, University of Warsaw, Ho\.{z}a 69, 00-681 Warsaw, Poland}
\author{P. Wojnar}
\affiliation{Institute of Physics, Polish Academy of Sciences, Lotnik\'{o}w 32/46, 02-668 Warsaw, Poland}
\author{P. Kossacki}
\affiliation{Institute of Experimental Physics, University of Warsaw, Ho\.{z}a 69, 00-681 Warsaw, Poland}

\date{\today}

\begin{abstract}

We present the photoluminescence excitation study of the self-assembled CdTe/ZnTe quantum dots doped with manganese ions. We demonstrate the identification method of spontaneously coupled quantum dots pairs containing single Mn$^{2+}$ ions. As the result of the coupling, the resonant absorption of the photon in one quantum dot is followed by the exciton transfer into a neighboring dot. It is shown that the Mn$^{2+}$ ion might be present in the absorbing, emitting or both quantum dots. The magnetic properties of the Mn$^{2+}$ spin are revealed by a characteristic sixfold splitting of the excitonic line. The statistics of the value of this splitting is analyzed for the large number of the dots and gives the information on the maximum density of the neutral exciton wave function. 
\end{abstract}


\maketitle

\section{Introduction}

The ultimate limit of the miniaturization of the magnetic memory is the device operating on a single magnetic ion. Such a limit opens a new area of physics of individual objects called solotronics \cite{flatte2011}. The semiconductor quantum dots (QDs) with single magnetic ions are one of the possible realizations of such an idea \cite{kobak2013}. So far the spectroscopic studies were done for the emission of single QDs containing Mn$^{2+}$ ions in CdTe \cite{besombes2004, leger2005prl, leger2006, legall2009, legall2011, besombes2005, goryca2009, goryca2010, legall2010, koperski2011} and GaAs based systems \cite{kudelski2007,krebs2009}. Several properties and functionalities were investigated. Among them, there were studies of spin orientation \cite{legall2009, goryca2009}, spin relaxation \cite{legall2010}, magnetic ion impact on the excitonic states \cite{glazov2007} and nonlinear response \cite{legall2011}. The II-VI system is particularly convenient for such studies because of the isoelectronic character of the Mn$^{2+}$ ion. Additionally, the CdTe/ZnTe self-organized dots exhibit a natural tendency to form clusters of QDs. It was shown that in usual MBE grown samples there is a possibility of finding QD pairs, between which the exciton is transferred within a few picoseconds \cite{kazimierczuk2009}. Such pairs are well detectable by photoluminescence excitation (PLE) measurements. The dot pairs were used to demonstrate efficient polarization conversion in a single dots pair as well as to increase efficiency of the optical orientation of a single Mn$^{2+}$ ion \cite{goryca2009}. 

Here we combine the two properties of the CdTe/ZnTe system: opportunity to form dot pairs with feasibility of incorporation of single magnetic ions in the dots. We demonstrate heterostructures in which all scenarios are possible: absorption and emission in two different nonmagnetic dots, absorption in the dot with a magnetic ion and emission in nonmagnetic one, absorption in the nonmagnetic dot and emission in the dot with a Mn$^{2+}$ ion and both absorption and emission in dots with Mn$^{2+}$ ions. Additionally we analyze the large statistics of the dots with Mn$^{2+}$ ions and show how the characteristic splitting of the neutral exciton line is related to the maximum of the neutral exciton wave function.

\section{Experimental method and sample growth}
The samples used in our study were grown by molecular beam epitaxy (MBE) technique. They contained self-assembled CdTe QDs embedded in ZnTe barrier. The formation of QDs was induced by the amorphous tellurium desorption method \cite{tinjod2003}. The Mn$^{2+}$ ions were introduced into the CdTe layer, from which the QDs were formed. One of the more important parts of the growth procedure was the correct calibration of the amount of deposited manganese which determined the probability of finding exactly one magnetic impurity in a given QD. The adjustment of concentration was done by the control of the temperature of the Knudsen effusion cell and by the time for which the cell was open. We note that the optimum concentration needed for effective formation of coupled QD pairs with a single Mn$^{2+}$ ion in the absorbing dot is higher than the concentration optimized for finding a singly-doped emitting QD. This observation is most likely related to the fact that on average the excitonic transition in an absorbing QD has relatively high energy and thus such QDs are characterized by smaller volume than typical QDs observed in photoluminescence studies.

The identification of coupled QDs pairs with magnetic ions was possible in the photoluminescence excitation measurements (PLE). They were performed at pumped liquid helium temperature equal to 1.7 K. A tunable dye laser was used for excitation within the energy  range from 2080 to 2200 meV. The power of the laser was stabilized by a liquid crystal modulator, controlled in a feedback loop with the signal from a photodiode taken just before the entrance to the cryostat. An immersion reflection microscope was attached directly to the sample. The width of the laser spot at the sample's surface did not exceed 1$\mu$m. The luminescence from the QDs was spectrally resolved by a monochromator with a CCD camera attached.

\section{Optical properties of quantum dots with a single Mn$^{2+}$ ion}

The presence of a single Mn$^{2+}$ ion in a QD determines the properties of the excitonic PL spectra. The exchange interaction between the magnetic impurity and the heavy-hole exciton lifts up the degeneracy of the X-Mn complex with respect to the projection of Mn$2^{2+}$ spin on the growth direction \cite{besombes2005}. The source of the excitonic anisotropy is the momentum of the heavy hole equal to $\pm 3/2$ \cite{leger2005}. The Mn$^{2+}$ ion has spin equal to $5/2$, therefore the PL spectrum shows characteristic sixfold splitting of the neutral exciton line \cite{besombes2004}. 

\begin{figure}
\includegraphics[width=8cm]{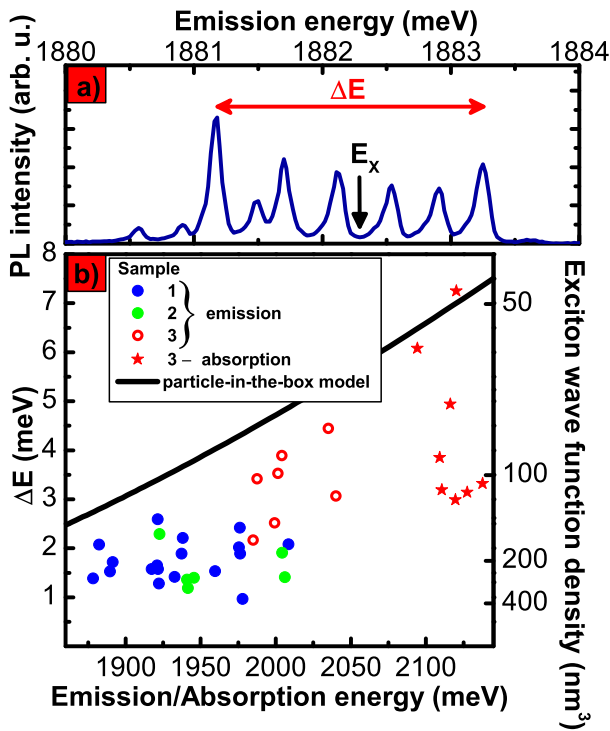}
\caption{(a) An exemplary PL spectrum of a neutral exciton confined in a single QD containing one Mn$^{2+}$ ion. (b) The statistical distribution of the value of the splitting induced by X-Mn exchange interaction as a function of emission energy for single or coupled QDs with a Mn$^{2+}$ ion in the emitting dot or absorption energy for coupled QDs with a Mn$^{2+}$ ion in the absorbing dot. The black line shows the results of calculations in the box model of a QD which reproduces the maximum of the wave function density for a given neutral exciton energy. \label{fig:stat}}
\end{figure}

An exemplary neutral exciton PL spectrum of a QD with a single Mn$^{2+}$ ion is presented in the figure \ref{fig:stat}(a). The dot was selected from the low energy tail of the broad emission band. The characteristic pattern related to different charge states was observed as described in Ref. \onlinecite{kudelski2005}. Here we focus on the neutral exciton. Additionally to the six main lines in the excitonic spectrum it is possible to observe weak lines corresponding to the brightened state of the dark exciton \cite{goryca2010}. This effect is the result of the combined effect of two factors: the X-Mn exchange interaction and the valence-band mixing \cite{leger2007}. The anisotropy of the QD's shape and the anisotropy induced by the in-plane strain are the main reasons for the presence of the light hole admixture in the hole ground state in a QD. The detailed description of the mechanism leading to the observation of the dark states can be found in Ref. \onlinecite{goryca2010}. Recent results obtained for CdTe QDs also indicate that there exists an in-plane radiative recombination channel of the dark exciton without external magnetic field even for nonmagnetic QDs \cite{smolenski2012}.

\begin{figure}
\vspace{0.2cm}
\includegraphics[width=8cm]{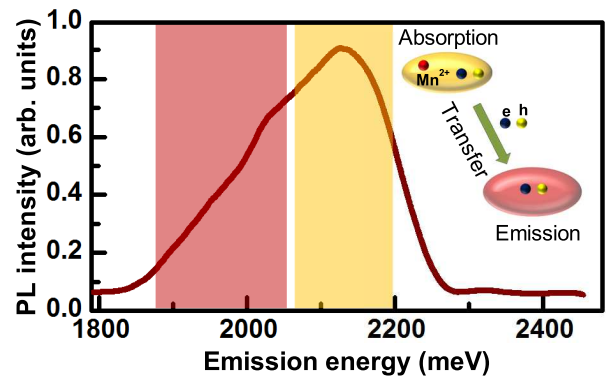}
\caption{The macro-photoluminescene (PL) spectrum of the CdTe QDs ensemble. In PLE measurements the excitation energy range is set around the maximum of the PL band. At the detection window energetic density of QDs is low enough to distinguish lines from single QDs in a micro PL experiment. The cartoon in the top-right part of the figure presents the exciton transfer mechanism between two coupled QDs. \label{makro}}
\end{figure}

The strength of the interaction between the magnetic impurity and the neutral exciton is characterized by the exchange constants $I_e$ (electron-Mn exchange integral) and $I_h$ (hole-Mn exchange integral). These quantities are the main factors that determine the sixtuplet splitting ($\Delta E$ in the figure \ref{fig:stat}(a)) of the bright excitonic line. The exchange splitting $\Delta E$ can be estimated by the following formula: $\Delta E = \frac{5}{2} \left( I_e - 3 I_h \right)$. This splitting is a good measure of the effectiveness of the incorporation of the Mn$^{2+}$ ion into a QD in terms of the overlap of the Mn$^{2+}$ ion wave function with the excitonic wave function. Figure \ref{fig:stat}(b) shows the statistics of the $\Delta E$ as a function of neutral exciton emission or absorption energies ($E_X$ in the figure \ref{fig:stat}(a)) for three different samples with various Mn$^{2+}$ ions concentration. The statistics was considerably expanded in respect to previous studies by including the absorption data. The dots used in the analysis were selected among those with well resolved sixtuplet and thus correspond to the cases with the location of the Mn$^{2+}$ ion in the vicinity of the center of the QD. Therefore the upper limit of the spread of the experimental points is most significant. The distribution shows the tendency to rise the value of the $\Delta E$ splitting with the increase of the emission or absorption energy. It follows the same trend for different samples of higher or lower Mn doping. The upper limit of the splitting for lower energy range equals to about 2.5 meV, while for higher energies the limit is equal to about 7.0 meV. We applied a simple model to estimate the exciton wave function density at the position of Mn$^{2+}$ ion. Assuming the same electron and hole densities resulting in $I_e/I_h=-3/4$ \cite{gaj1979} and  neglecting the e-h interaction we compare the exchange constants $I_e$ and $I_h$ with the (Cd,Mn)Te material constants $N_0 \alpha=0.22$ eV  and $N_0 \beta=-0.88$ eV \cite{gaj1979}. By calculating the energy of h-Mn complex in two different hole wave function basis (using either the effective spin $j=\pm3/2$ or the p-type orbital wave function combined with the electron spin $\sigma=\pm1/2$) we establish the excion wave function density as (analogous calculations for the e-Mn complex lead to the formula with electronic parameters):
\begin{equation}
\frac{1}{\left|\Psi\left(0\right)\right|^2}=\frac{N_0 \beta}{6 I_h} V_0 = \frac{N_0 \alpha}{2 I_e} V_0
\end{equation}
where $\left|\Psi\left(0\right)\right|^2$ is the excitonic wave function density at the position of Mn$^{2+}$ ion and the $V_0$ is the volume of the CdTe material elementary cell. The obtained value of the inverse of the density for the QDs with a Mn$^{2+}$ ion was found in the range from 50 nm$^3$ up to 400 nm$^3$. This value is surprisingly small and leads to the conclusion of strong localization of the exciton wave function in the dot. We compare it with a trend calculated in a simple particle in a box model which gives an estimation of the dependence of the wave function density on the neutral exciton energy. We assumed the cuboid infinite well potential confining a noninteracting electron and hole in the effective mass approximation with the CdTe material parameters \cite{lesidang1982, fishman1995, smolenski2011}. The curve presented in the figure \ref{fig:stat}(b) was obtained by varying the QD lateral width while keeping constant the QD height. This simple model describes the experimental data with good agreement for reasonable sizes of the box: hight equal to 3.5 nm and the width ranging from 5 nm to 20 nm.

\section{PLE spectra of quantum dots pairs}

To extend the studies of the single impurity in a QD to the system of coupled QDs we searched for sharp resonances in PLE spectra. The highest probability of finding the dot pair suitable for our studies was observed for exciting energy close to the maximum of the macro-PL band of QD ensemble and detection energy range located at the low energy tail. Both ranges are marked in figure \ref{makro}.  The figure also shows a macro PL spectrum of a typical sample.  The highest spectral density of QDs in the excitation range ensures high probability of tuning the laser in resonance with the absorbing dot. The detection window assures that the spectral density of emitting QDs is low enough to distinguish isolated lines from single QDs in micro PL spectrum.

\begin{figure}
\includegraphics[width=8cm]{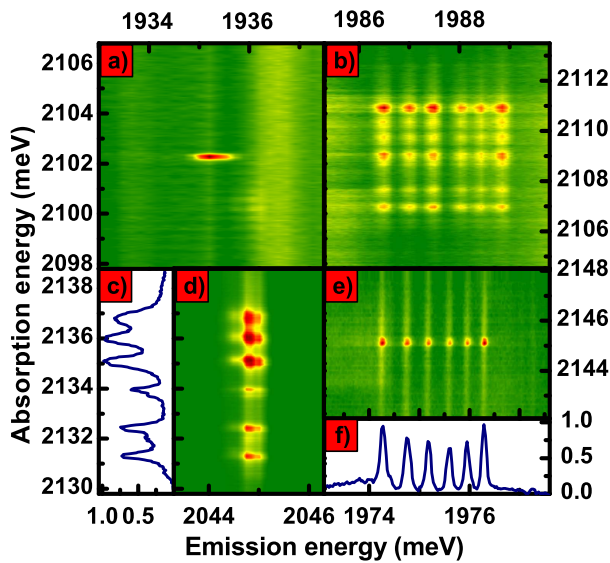}
\caption{PLE maps reveal spontaneous coupling between (a) nonmagnetic QDs, QDs with Mn$^{2+}$ ion in (c) absorbing, (d) emitting and (b) both QDs. Figures (e) and (f) show cross sections taken at resonant energies (2044.8 meV and 2145.1 meV, respectively), which correspond to the absorption and emission spectrum of a QD with a single Mn$^{2+}$ ion.\label{ple}}
\end{figure}

The results of the PLE measurements revealed the occurrence of resonant emission for specific excitation energies. It has been shown that observed resonances arise from the efficient exciton transfer between QDs forming a coupled pair \cite{kazimierczuk2009, kazimierczuk2010, koperski2011_2}. An example of a resonance for a nonmagnetic QDs pair is presented in figure \ref{ple}(a). In this case, the exciton is resonantly created at 2102 meV and transferred to a neighboring dot, where recombination occurs at 1935 meV (see top-right part of figure \ref{makro}). 

The systematic studies of large statistics of the PLE spectra show that the Mn$^{2+}$ ion might be incorporated in both QDs forming a pair. Thus we may observe three different configurations: the single Mn$^{2+}$ ion confined in the emitting, absorbing or both QDs. The first of these three cases was studied before \cite{goryca2009} and  opens a possibility to demonstrate the optical writing and read-out of the information stored in the Mn$^{2+}$ ion spin state. It was achieved by exploiting the orientation of the Mn$^{2+}$ ion by injection of polarized photocarriers into the emitting QD. The typical PLE map with the sixfold spitting of the resonance with respect to the emission energy is presented in figure \ref{ple}(e).

The qualitatively new system we analyze here consists of a coupled QDs with a Mn$^{2+}$ ion in the absorbing dot. For this system the characteristic sixfold splitting of the resonance occurs with respect to the absorption energy (see figure \ref{ple}(d)). A series of samples with different concentration of magnetic ions was grown and then characterized by the PLE measurement. It turned out that higher concentration of magnetic impurities significantly increased the probability of finding exactly one Mn$^{2+}$ ion in the absorbing dot. This effect can be explained if we assume that the size of the absorbing dots is smaller. Such an assumption is consistent with the fundamental requirement for the coupling of QDs pair, which is the greater energy of the neutral exciton in the absorbing dot. The smaller size of the absorbing dot would lead to greater spatial confinement of the neutral exciton and therefore cause the increase of its energy. Another feature that would justify such an assumption can be found in the statistics of the exchange splitting as the function of the absorption energy. The mean value of the splitting energy is equal to about 4 meV, that is two times more than in the case of the emitting dots. This value of the splitting differs strongly between various dots, reaching in extreme cases 7 meV.   

The third case when the single magnetic ions occupy both absorbing and emitting QD was also observed. For this system a 6$\times$6 pattern of resonances occurs in the PLE map (figure \ref{ple}(b)). The presented map reveals no fingerprint of the antiferromagnetic coupling between the two Mn$^{2+}$ ions in the two neighboring dots as we do not observe the enhancement of resonant emission for states corresponding to the antiparallel alignment of the Mn$^{2+}$ ions spins. The observed resonances scheme suggests that both ions independently probe the excitonic wave function because the cross-section emission or absorption spectrum to a small degree depends on the choice of the row or column in the grid of resonances. 

\begin{figure}
\includegraphics[width=8cm]{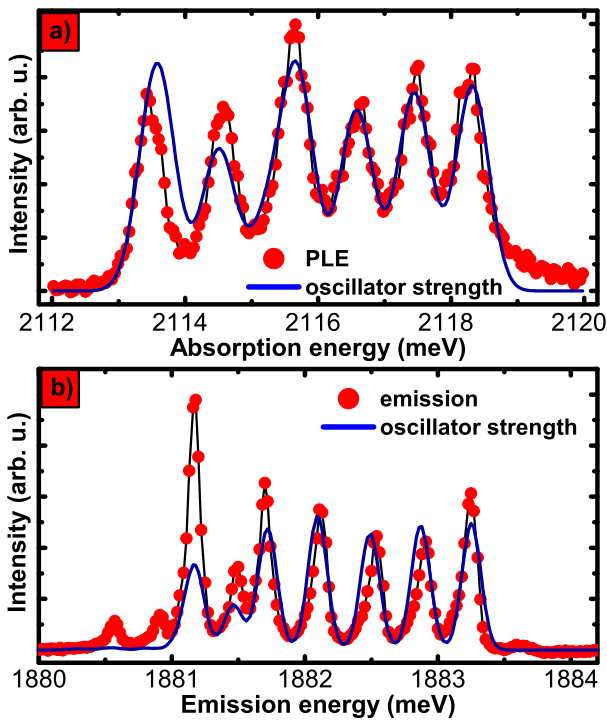}
\caption{(a) PLE and (b) emission spectrum of two different QDs with a single Mn$^{2+}$ ion compared with the numerical simulation results, which reproduce the oscillator strengths of excitonic transitions.}
\label{fit}
\end{figure}

Figures \ref{ple}(c) and \ref{ple}(f) present the PLE and emission spectra of a CdTe QD with a single Mn$^{2+}$ ion respectively, taken as the cross sections of the PLE maps at the resonant energies. A closer look at both spectra leads to the conclusion that analysis of the PLE has significant advantages over the PL data. In the PLE spectrum the lines intensities differ strongly among themselves while in emission spectra the intensities remain much more uniform. 
This reflects different origin of the line intensity in the absorption and in the emission spectrum. In the emission, it is governed mostly by the dynamics of the excitation and uniform intensities correspond to the same probability of finding the Mn$^{2+}$ ion in one of the states with different Mn$^{2+}$ ion spin projections. The line intensities in the PLE spectrum with a good approximation correspond to the values of the oscillator strengths of the excitonic transitions. These values are strongly affected by the exciton anisotropy and the light-heavy hole mixing, but they might be easily described by the excitonic model (ref. \onlinecite{goryca2010}). In the figure \ref{fit}(a) the PLE spectrum of a selected QD with a single Mn$^{2+}$ ion is compared with a numerical simulation. The theoretical model describes the pattern of intensities well. The slight discrepancies may arise from the fluctuations of the energy and effective power density of the dye laser used for the PLE measurements. On the other side, the PL spectra are mostly determined by the relaxation dynamics. The striking signature in figure 4(b) is the strong enhancement of the ''dark'' excitons intensity on the low energy side. They are well visible in spite of much smaller oscillator strength and long lifetime. Such difference is clear when comparing the data with the simulation. This analysis shows that the system of coupled QDs with Mn$^{2+}$ ion in absorbing dot can be applied to directly determine the oscillator strengths of excitonic transitions. Here one should note that another characteristic feature of the PLE spectrum is a quite large width of the lines. It results mostly from the homogeneous broadening related to the short lifetime of the exciton in the absorbing dot.

\section{Summary}

In summary, we have shown that in CdTe/ZnTe self-organized system of QDs it is possible to introduce single magnetic ions into pairs of coupled dots. All possible combinations of the location of the Mn$^{2+}$ ion were demonstrated: absorption and emission in two different nonmagnetic dots, absorption in the dot with magnetic ion and emission in nonmagnetic one, absorption in nonmagnetic and emission in dot with Mn$^{2+}$ ion, and both absorption and emission from dots with Mn$^{2+}$ ions. We used large statistics of dots to analyze the density of the exciton wave function in the center of the QD. The analysis of the extended statistics of the QDs with the Mn$^{2+}$ ions gives the access to the variation of the maximum density of the neutral exciton wave function throughout the population of the dots. We have also shown that the oscillator strength of the excitonic transitions might be directly studied by the analysis of the PLE spectrum.

This work was supported by the Polish Ministry of Science and Higher Education in years 2012--2016 as research grants ''Diamentowy Grant'', ''Preludium'' and ''Iuventus'', by European Project MTKD-CT-2005-029671, NCBiR project LIDER, NCN projects DEC-2011/01/B/ST3/02406 and DEC-2011/02/A/ST3/00131 and FNP subsidy ''Mistrz''. Experiments were carried out with the use of CePT, CeZaMat and NLTK infrastructures financed by the European Union - the European Regional Development Fund within the Operational Programme "Innovative economy" for 2007--2013.

\end{document}